\documentclass[11pt,twoside]{article}
\usepackage{asp2010}

\resetcounters

\begin{document}

\title{Type Ia supernovae and the DD scenario}
\author{J. Isern$^{1,2}$, E. Garc{\'\i}a--Berro$^{2,3}$, and P. Lor\'en--Aguilar$^{2,3}$}
\affil{$^1$Institut de Ci\`encies de l'Espai (CSIC), 
           Facultat de Ci\`encies, Campus UAB, 
           Torre C5-parell, 
           08193 Bellaterra, Spain}
\affil{$^2$Institute for Space  Studies of Catalonia,
           c/Gran Capit\`a 2--4, Edif. Nexus 104,   
           08034  Barcelona,  Spain}
\affil{$^3$Departament de F\'\i sica Aplicada,
           Universitat Polit\`ecnica de Catalunya,
           c/Esteve Terrades 5, 
           08860 Castelldefels, Spain}

\begin{abstract}
Type Ia supernovae are thought  to be the outcome of the thermonuclear
explosion of  a white  dwarf in a  close binary system.   Two possible
scenarios,  not  necessarily incompatible,  have  been advanced.   One
assumes  a  white dwarf  that  accretes  matter  from a  nondegenerate
companion  (the single  degenerate  scenario), the  other assumes  two
white  dwarfs  that  merge  as   a  consequence  of  the  emission  of
gravitational waves (the double  degenerate scenario).  The delay time
distribution of  star formation bursts  strongly suggests that  the DD
scenario   should   be  responsible  of  the  late  time    explosions 
\citep{TEA08} , but this contradicts the common   wisdom   that  the 
outcome of the merging of  two  white  dwarfs  is an accretion induced 
collapse to  a neutron star.  In  this contribution we  review some of 
the most  controversial issues of this problem.
\end{abstract}

\section{Introduction}

Supernovae are characterized by a  sudden rise of their luminosity, by
a steep decline after maximum light that lasts several weeks, followed
by  an exponential  decline that  can  last several  years. The  total
energy involved output turns out to be $\sim 10^{51}$~erg. Such amount
of energy  can only  be obtained in  two ways, from  the gravitational
collapse of  an electron degenerate core  to form a neutron  star or a
black hole  or from the thermonuclear incineration  of a carbon-oxygen
degenerate core.

From  the  spectroscopic  point   of  view,  Type  Ia  supernovae  are
characterized  by  the absence  of  hydrogen  and  the presence  of  a
prominent  Si~II  line  at  maximum  light  (in  fact  these  are  the
characteristics that define the  class). From the photometric point of
view  they experience  a sudden  rise  to maximum  of light,  $\langle
M_{\rm B}\rangle \sim -19^{\rm mag}$,  in 20 days or less, followed by
a sudden decline of $ \sim 3^{\rm mag}$ in 30 days, and an exponential
tail with  a characteristic time  $ \sim 70$  days. From the  point of
view of  the galaxies hosting supernovae, it  was immediately realized
that SNIa  occur in all  galaxy types in  contrast with all  the other
supernova  types  that  only  appear  in  galaxies  with  recent  star
formation activity.   It is interesting  to notice here that  the SNIa
rate   correlates  with  the   star  formation   rate  \citep{sull06},
indicating that  the progenitors  must be able  to produce  prompt and
delayed  explosions.   All  these  arguments immediately  led  to  the
conclusion that SNIa  were caused by the thermonuclear  explosion of a
carbon-oxygen  white dwarf  near  the Chandrasekhar  mass  in a  close
binary  system,   an  idea   strongly  supported  by   the  remarkable
spectro-photometric homogeneity  displayed by this  kind of supernovae
\citep{cado85,fili97}.

However, despite the homogeneity  of Type Ia supernovae, when observed
in detail some  differences have appeared over the  years. It is known
there is  a group of  SNIa with light  curves showing very  bright and
broad peaks,  the SN1991T  class, that represents  the 9\% of  all the
events. There  is another  group with very  dim and narrow  peaks that
lack of the  characteristic secondary peak, the SN1991bg  class , that
represent the 21\% of all the  events. To these categories it has been
recently added a  new one that contains very  peculiar supernovae, the
SN2002cx class,  that represents  the $\sim 4$\%  of the  total. These
supernovae are characterized by a high ionization spectral features in
the pre-maximum, like the SN1991T  class, and by a very low luminosity
and  the lack  of secondary  maximum,  like the  SN1991bg class.   The
remaining ones, which amount  to $\sim 66\%$, present normal behaviors
and are  known as {\sl  Branch-normal}. However, even the  normal ones
are  not completely  homogeneous  and show  different luminosities  at
maximum and light curves with different decline rates \citep{li11b}.

Despite the substantial advances made during the last years, the basic
question of which systems explode has not been satisfactorily answered
yet.   There is  a  wide consensus  that  the progenitors  have to  be
composed by a  carbon-oxygen white dwarf and a  close enough companion
able  to  provide the  mass  necessary  to  trigger the  thermonuclear
instability.   Several evolutionary paths  leading to  the instability
have been  identified \citep{post06}. The systems able  to explode can
be  classified  according  to  the  nature of  the  donor  (normal  or
degenerate stars, also known as  SD or single degenerate scenarios and
DD or double degenerate scenarios) and the composition of the accreted
matter (hydrogen, helium or a mixture of carbon and oxygen).

The first  systems identified  as potential supernova  progenitor were
those formed  by a white dwarf that  accretes hydrogen \citep{whel73}.
There  are  many  types   of  these  systems:  cataclysmic  variables,
classical novae, recurrent novae,  symbiotic stars and supersoft X-ray
sources.   If  the accretion  rate  is  smaller  than $\sim  10^{-9}\,
M_\odot$~yr$^{-1}$,  the   accreted  matter  becomes   degenerate  and
experiences   a  strong   flash  that   can  be   identified   with  a
nova. However,  such events not only  eject all the  accreted mass but
also erode  the white dwarf  preventing it from reaching  the critical
mass. The accreted  mass can only be retained if  $M_{\rm WD} > 1.35\,
M_\odot$, but white dwarfs with such initial masses are made of oxygen
and neon \citep{GPGB01,GP03} and cannot  explode as a SNIa. Thus, only
white dwarfs  that have experienced  a previous accretion  episode can
reach the Chandrasekhar's mass  at such rates. For intermediate rates,
$10^{-9}  \le   \dot{M}_{\rm  H}\,  (M_\odot/{\rm   yr})  \le  5\times
10^{-7}$,  hydrogen  burns  steadily  or through  mild  flashes  which
accumulate a  helium buffer on top  of the carbon-oxygen  core. If the
accretion rate is high enough,  the freshly formed helium is converted
into carbon and oxygen through  weak flashes or steady burning and the
white  dwarf  can approach  to  the  Chandrasekhar  mass. But  if  the
accretion rate is roughly in  the range $10^{-9} \le \dot{M}_{\rm H}\,
(M_\odot/{\rm  yr}) \le  5\times 10^{-8}$,  the helium  layer explodes
under degenerate conditions  and can trigger the explosion  of all the
star \citep{nomo82b}.

One of  the problems  posed by the  SD scenario  is that to  avoid the
constraint  imposed by novae,  the accretion  rates have  initially to
proceed at a high rate and this makes white dwarfs hot and potentially
detectable as X-ray sources. If it  is assumed that all SNIa come from
supersoft X-ray sources, the number necessary to sustain the estimated
rate of  supernovae in  M31 or  the Milky Way  is $\sim  1\,000$ while
current surveys have  only detected between $\sim 10$  and 100. In the
case of elliptical galaxies the problem seems to be even worst. At the
first   glance   it  seems   that   this   argument   favors  the   DD
scenario. However,  to form  a DD the  system must evolve  through two
common envelope stages  and this means that they  will appear for some
time,  $> 10^6$~yr,  as symbiotic  stars  with a  white dwarf  burning
hydrogen at the  surface. It is therefore clear  that the problem lies
on the  observational properties of these  high-rate accreting sources
\citep{dist10b}.

A possible way to overcome this problem could consist on the formation
of heavy winds  able to hide the nuclear burning  white dwarf, as well
as to  consider variable accretion rates able  to produce successively
symbiotic stars, supersoft X-ray sources  and, when the white dwarf is
massive enough,  recurrent novae. In any  case, plausible evolutionary
paths have  been advanced \citep{hach08}. Minor  possible objections to
this  scenario come from  the non-detection  of the  hydrogen stripped
from  the companion  by the  explosion  and the  non-detection of  the
surviving star in the case of recent events in the Milky Way.

Close  enough  binaries  formed  by two  intermediate-mass  stars  can
experience two episodes of common envelope evolution and form a double
degenerate  system  made  of  two  carbon-oxygen  white  dwarfs.   The
emission of gravitational waves further  reduces the orbit and, if the
separation  of both  white dwarfs  is not  too large,  they eventually
merge in  less than  a Hubble time  \citep{webb84, iben85}.   The main
advantages of the DD scenario are that there are several known systems
able to merge in a short time, although their mass is smaller than the
critical  mass,  and that  the  time  distribution  of the  persistent
component  of SNIa  in galaxy  clusters  is only  compatible with  the
merging of two  white dwarfs \citep{maoz10}. The main  problem of this
scenario  is that  if  the accretion  rate  is spherically  symmetric,
$\dot{M} \ge  2.7 \times 10^{-6}\, M_\odot$~yr$^{-1}$  and the entropy
of  the infalling  matter is  neglected, the  white dwarf  ignites off
center  and becomes  an  oxygen-neon white  dwarf \citep{nomo85}  that
eventually collapses to a  neutron star \citep{nomo91}. Probably, this
approach is extremely simple since the accreted mater has also angular
momentum that it is incorporated  to the white dwarf.  Therefore it is
obvious  that the  coupling between  the disk  and the  rotating white
dwarf is a critical issue \citep{pier03b,pier03a,saio04}.

There are  two scenarios in which  a white dwarf  can directly accrete
helium. One  is a  double degenerate with  a secondary made  of helium
that merge  as a consequence  of the emission of  gravitational waves,
the other is a single degenerate  scenario in which the secondary is a
non-degenerate helium star and the  mass transfer is powered by helium
burning. One-dimensional  models indicate that helium  can ignite just
above  the  base  of  the  freshly accreted  mantle  under  degenerate
conditions.  The high  flammability of  helium together  with  the low
density of  the envelope  induces the formation  of a  detonation that
incinerates the  envelope and triggers the  thermonuclear explosion of
all the  accretor despite the fact  that its mass is  smaller than the
Chandrasekhar's mass \citep{nomo82b}.

Sub-Chandrasekhar models are not the favorites to account for the bulk
of supernovae. The reason is that they predict the existence of a very
fast moving  layer made of $^{56}$Ni  and $^4$He that  is not observed
\citep{hoef96}.   However,   if  it  were  possible   to  neglect  the
contribution  of  this  outer  layer, these  explosions  could  nearly
reproduce  the  gross  features  of  SNIa  explosions  with  a  single
parameter, the initial mass of  the white dwarf \citep{sim10}. In this
sense, it is important to realize that the process of formation of the
mantle is quite complex and  far from the simple description of matter
accreted with spherical symmetry, at a constant rate and with the same
entropy  as  the envelope,  opening  the  possibility  to trigger  the
detonation of the white dwarf  without incinerating the outer He layer
\citep{guil10}.

\section{The fate of DD mergers}

\begin{figure} 
\vspace{8cm}
\includegraphics{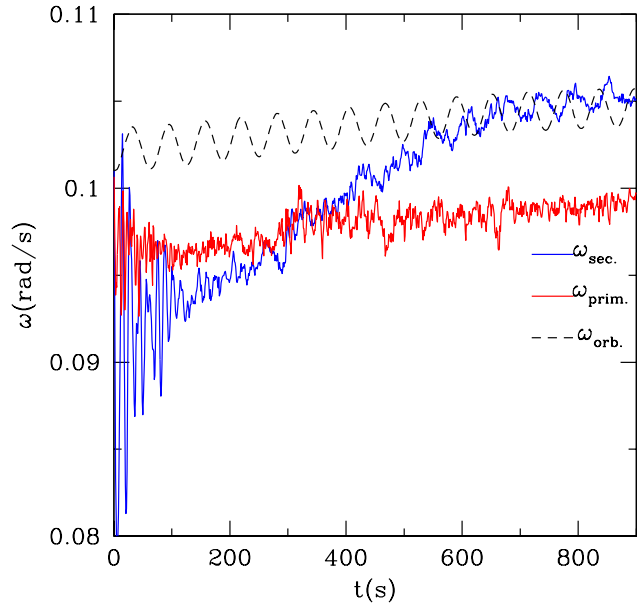} 
\caption{Evolution of  the orbital and rotation velocities  of the two
  white dwarfs before merging.}
\label{fig1}
\end{figure}

The temporal  evolution of  the merging of  two white dwarfs  has been
computed     using     SPH     techniques    by     several     groups
\citep{benz90,rasi94,segretain97,guer04,ross07,lore09}.     A   common
feature of all these simulations is that the secondary is destroyed in
a dynamical timescale,  i.e. a few orbital periods,  after filling the
Roche  lobe  and forms  a  thick and  hot  accretion  disk around  the
primary. The impact  of the disrupted secondary on  the primary is not
able  to induce  a prompt  ignition \cite{guer04}  and a  rotating hot
corona surrounded  by a thick  disk formed. Because of  the timescales
involved, a  detailed self-consistent  simulation of the  evolution of
such   a  configuration   (corona   and  disk)   has   not  yet   been
performed.  However, by mapping  the 3D  results into  1D simulations,
\cite{yoon07} were able to find the conditions to avoid the off-center
ignition. They  found that the  interface do not reaches  the ignition
temperature  when  quasi-static  equilibrium  is  achieved,  that  the
timescale  for neutrino  cooling is  shorter than  the time  scale for
angular momentum dissipation, and finally that the mass accretion rate
is $\le 5\times 10^{-6},\,10^{-5}\, M_\odot$~yr$^{-1}$.

These results  were challenged  by \cite{dsou06} and  \cite{motl07} on
the  basis of simulations  performed using  grid techniques.  The main
difference was that in grid  based methods the donor was not destroyed
on a  dynamical timescale but  suffered mass transfer  episodes during
many orbital periods.  Therefore, if the mass transfer  is stable, the
accretion rate would be determined  by the rate at which gravitational
waves  remove  angular  momentum.    The  problem  was  reexamined  by
\cite{dan11}, who have  shown that the origin of  the discrepancies is
at the  initial conditions  imposed to the  SPH simulations.  In these
simulations, the  Roche limit was was  very deep in  the secondary and
the  initial  mass transfer  very  large.  When  the reliable  initial
conditions are adopted, carefully allowing the stars to relax to their
equilibrium configuration  before the beginning of  the mass transfer,
the binary can survive for hundreds of orbital periods \citep{dan11}.

\begin{figure} 
\centering
 \includegraphics[width=0.8\hsize]{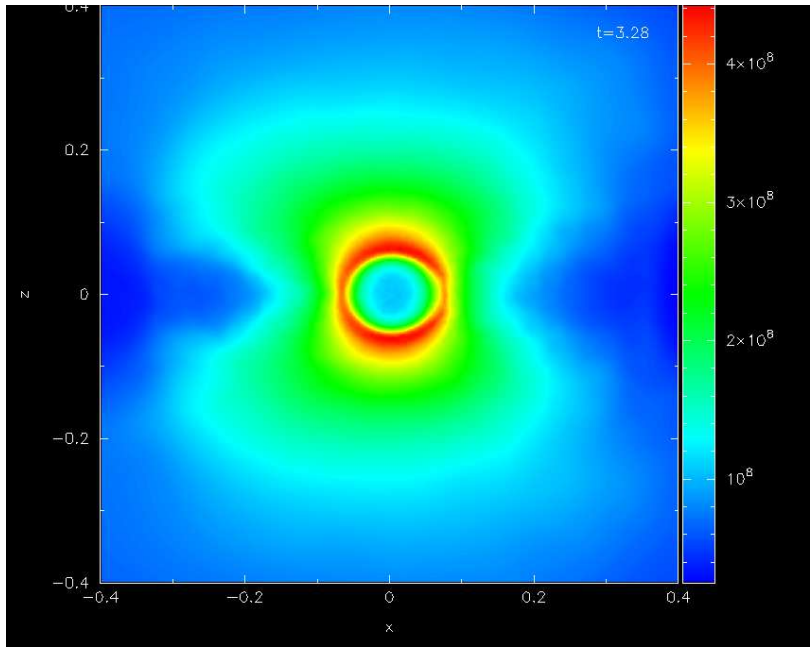}
 \caption{Temperature  distribution after merging.  The presence  of a
   hot corona surrounded by a thick Keplerian disk is clearly seen.}
 \label{fig2}
\end{figure}

We have performed a new simulation of the merging of two white dwarfs,
of 0.8 and  $0.6\, M_\odot$ respectively, made of  a realistic mixture
of carbon and oxygen.  The  procedure was the same as in \cite{lore09}
except that  the initial conditions  were obtained more  carefully. In
this case we introduced an  artificial braking term and we allowed the
two   stars   to   approach   to  their   equilibrium   configuration.
Figure~\ref{fig1} displays the evolution of the orbital and rotational
velocities after the first episode  of braking that shows the tendency
to synchronize.   Our results  clearly confirm those  of \cite{motl07}
and  \cite{dan11}, namely  that  the  merging does  not  proceed on  a
dynamic timescale but it can  take many orbits to occur. However, once
started, the accretion rate tends to increase rapidly until it becomes
catastrophic, leading to the formation of a hot corona surrounded by a
thick disc  around the primary (see  figure~\ref{fig2}). However, this
behavior is the result of  preliminary calculations, and can be due to
the low mass  resolution, $\sim 5\times 10^{-6}\, M_\odot$,  so it has
to be taken with some care.

\begin{figure} 
\vspace{8cm}
\includegraphics{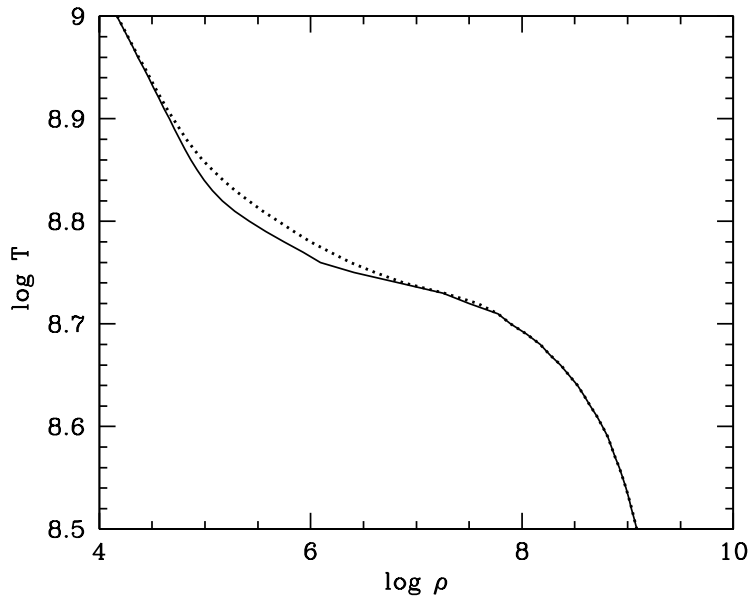} 
\caption{Ignition line for carbon rich matter. The dotted line includes axions}
\label{fig3}
\end{figure}

Since the  masses of  carbon-oxygen white dwarfs  are not  larger than
$\sim  1.1\,  M_\odot$  \citep{GB97,GB99},  the  primary  is,  at  the
beginning, always  far from the Chandrasekhar's  mass and consequently
it will  not be able  to reach the  critical value before the  bulk of
mass transfer  sets in. Thus, it  is hard to see,  if our calculations
are  correct, how  to avoid  the  episode of  high mass-transfer  that
follows  the total  disruption  of the  secondary  and the  subsequent
off-center  ignition  of  carbon  followed  by  an  accretion  induced
collapse.

\section{Discussion and conclusions}

There are,  in principle,  three ways to  to account for  the observed
\citep{maoz10}  temporal distribution of  supernova eruptions:  i) for
some physical  reason the ignition  is prevented, ii) the  rotation of
the primary stabilizes the accretion  rate to a value of $\sim 10^{-7}
\,  M_\odot$~yr$^{-1}$, eventually leading  to a  supernova explosion,
and  iii)  the  responsible  of  the  explosion is  the  merger  of  a
carbon-oxygen white dwarf and a helium white dwarf. We examine them in
the following.

The reason why the first point has to be considered is twofold. On one
hand,  the cross  section  of the  $^{12}$C+$^{12}$C  reaction at  low
energies is not well known.   Firstly, there are some claims about the
existence of low energy resonances,  which would favor the ignition of
carbon (Bravo et  al., in preparation), but also, there  is a claim of
the existence of a hindrance that could prevent the ignition of carbon
at low  temperatures \citep{gasq07}.   On the other,  the size  of the
network  of reactions usually  adopted to  describe carbon  burning is
usually minimal, so carbon burning is treated in a very simplified way
\citep{fost10}.  Another possibility comes  from the claim that axions
exist and could  be responsible of the suspected  anomalous cooling of
white  dwarfs  \citep{iser08,iser10}.  If  this  were  the  case,  the
ignition  line would be  modified just  in the  region of  the density
temperature plane where the  off-center ignition of carbon occurs (see
figure~\ref{fig3}) and the velocity  of propagation of the flame would
also change. The impact of such effect has not been yet elucidated.

The second  point relies  on the  fact that if  the accretion  rate is
initially high, near the  Eddington limit, as expected, thermal energy
has  no time  to  diffuse from  the  outer layers  and produces  their
expansion. Meanwhile, the angular velocity of the primary increases as
a consequence  of the accreted  angular momentum until  the equatorial
layers reach  the critical  value where centrifugal  and gravitational
forces   become  equal   and   no   matter  can   be   added  to   the
configuration. This allows the star to redistribute the thermal energy
and  the angular  momentum  and re-start  the  accretion process  once
more. The  net result is that  the accretion rate is  regulated by the
heat  transfer time  \citep{pier03a}. As  the primary  heats  up, this
mechanism  becomes  less and  less  efficient  and  stops well  before
reaching  the  critical  mass.   If  the  primary  adopts  a  triaxial
configuration  as a  consequence  of rotation,  it  can loose  angular
momentum by  the emission  of gravitational waves  and the  process of
accretion  can   re-start  once  more  until  the   star  reaches  the
Chandrasekhar's mass  and explodes as a SNIa  \cite{pier03b}. The main
difficulty of this scenario is how  to couple the primary and the disk
to avoid mass losses.

The third possibility to obtain delayed explosions is the merging of a
helium and a  carbon-oxygen white dwarf. In this  case it is necessary
to prevent  the formation  of large envelopes  of helium to  avoid the
presence  of large amounts  of $^{56}$Ni  in the  outer layers  of the
supernova debris.  This could be  achieved by focusing the  shock wave
induced  by the detonation  of small  amounts of  helium in  the outer
layers of the primary \citep{fink10,guil10}.

In  conclusion, there  are strong  observational  evidences suggesting
that  the merging  of two  white dwarfs  could produce  some supernova
events, specially the most delayed  ones. The main difficulty to prove
that  this is  the case  is the  extreme difficulty  of  modelling the
long-term behavior of the disk and the primary.

\acknowledgements This  work has been  supported by the  MICINN grants
AYA08-1839/ESP and AYA2008-04211-C02-01,  by the ESF EUROCORES Program
EuroGENESIS  (MICINN grants  EUI2009-04167 and  04170), by  the grants
2009SGR1002 and 2009SGR315 of the Generalitat de Catalunya, and by the
European Union FEDER funds.

\bibliography{Isern_J}

\hyphenation{Post-Script Sprin-ger}
\begin{thebibliography}{}
\expandafter\ifx\csname natexlab\endcsname\relax\def\natexlab#1{#1}\fi
\expandafter\ifx\csname url\endcsname\relax
  \def\url#1{\texttt{#1}}\fi
\expandafter\ifx\csname urlprefix\endcsname\relax\def\urlprefix{URL }\fi
\providecommand{\eprint}[2][]{\url{#2}}

\bibitem[{{Benz} et~al.(1990){Benz}, {Cameron}, {Press}, \& {Bowers}}]{benz90}
{Benz}, W., {Cameron}, A.~G.~W., {Press}, W.~H., \& {Bowers}, R.~L. 1990, \apj,
  348, 647

\bibitem[{{Cadonau} et~al.(1985){Cadonau}, {Tammann}, \& {Sandage}}]{cado85}
{Cadonau}, R., {Tammann}, G.~A., \& {Sandage}, A. 1985, in Supernovae as
  Distance Indicators, edited by {N.~Bartel}, vol. 224 of Lecture Notes in
  Physics, Berlin Springer Verlag, 151

\bibitem[{{Dan} et~al.(2011){Dan}, {Rosswog}, {Guillochon}, \&
  {Ramirez-Ruiz}}]{dan11}
{Dan}, M., {Rosswog}, S., {Guillochon}, J., \& {Ramirez-Ruiz}, E. 2011, \apj,
  737, 89

\bibitem[{{Di Stefano}(2010)}]{dist10b}
{Di Stefano}, R. 2010, \apj, 719, 474

\bibitem[{{D'Souza} et~al.(2006){D'Souza}, {Motl}, {Tohline}, \&
  {Frank}}]{dsou06}
{D'Souza}, M.~C.~R., {Motl}, P.~M., {Tohline}, J.~E., \& {Frank}, J. 2006,
  \apj, 643, 381

\bibitem[{{Filippenko}(1997)}]{fili97}
{Filippenko}, A.~V. 1997, \araa, 35, 309

\bibitem[{{Fink} et~al.(2010){Fink}, {R{\"o}pke}, {Hillebrandt}, {Seitenzahl},
  {Sim}, \& {Kromer}}]{fink10}
{Fink}, M., {R{\"o}pke}, F.~K., {Hillebrandt}, W., {Seitenzahl}, I.~R., {Sim},
  S.~A., \& {Kromer}, M. 2010, \aap, 514, A53+

\bibitem[{{F{\"o}rster} et~al.(2010){F{\"o}rster}, {Lesaffre}, \&
  {Podsiadlowski}}]{fost10}
{F{\"o}rster}, F., {Lesaffre}, P., \& {Podsiadlowski}, P. 2010, \apjs, 190, 334

\bibitem[{{Garc{\'{\i}}a--Berro} et~al.(1997){Garc{\'{\i}}a--Berro}, {Ritossa},
  \& {Iben}}]{GB97}
{Garc{\'{\i}}a--Berro}, E., {Ritossa}, C., \& {Iben}, I., Jr. 1997, \apj, 485,
  765

\bibitem[{{Gasques} et~al.(2007){Gasques}, {Brown}, {Chieffi}, {Jiang},
  {Limongi}, {Rolfs}, {Wiescher}, \& {Yakovlev}}]{gasq07}
{Gasques}, L.~R., {Brown}, E.~F., {Chieffi}, A., {Jiang}, C.~L., {Limongi}, M.,
  {Rolfs}, C., {Wiescher}, M., \& {Yakovlev}, D.~G. 2007, \prc, 76, 035802

\bibitem[{{Gil-Pons} \& {Garc{\'{\i}}a--Berro}(2001)}]{GPGB01}
{Gil-Pons}, P., \& {Garc{\'{\i}}a--Berro}, E. 2001, \aap, 375, 87

\bibitem[{{Gil-Pons} et~al.(2003){Gil-Pons}, {Garc{\'{\i}}a--Berro},
  {Jos{\'e}}, {Hernanz}, \& {Truran}}]{GP03}
{Gil-Pons}, P., {Garc{\'{\i}}a--Berro}, E., {Jos{\'e}}, J., {Hernanz}, M., \&
  {Truran}, J.~W. 2003, \aap, 407, 1021

\bibitem[{{Guerrero} et~al.(2004){Guerrero}, {Garc{\'{\i}}a--Berro}, \&
  {Isern}}]{guer04}
{Guerrero}, J., {Garc{\'{\i}}a--Berro}, E., \& {Isern}, J. 2004, \aap, 413, 257

\bibitem[{{Guillochon} et~al.(2010){Guillochon}, {Dan}, {Ramirez-Ruiz}, \&
  {Rosswog}}]{guil10}
{Guillochon}, J., {Dan}, M., {Ramirez-Ruiz}, E., \& {Rosswog}, S. 2010, \apjl,
  709, L64

\bibitem[{{Hachisu} et~al.(2008){Hachisu}, {Kato}, \& {Nomoto}}]{hach08}
{Hachisu}, I., {Kato}, M., \& {Nomoto}, K. 2008, \apj, 679, 1390

\bibitem[{{Hoeflich} et~al.(1996){Hoeflich}, {Khokhlov}, {Wheeler}, {Phillips},
  {Suntzeff}, \& {Hamuy}}]{hoef96}
{Hoeflich}, P., {Khokhlov}, A., {Wheeler}, J.~C., {Phillips}, M.~M.,
  {Suntzeff}, N.~B., \& {Hamuy}, M. 1996, \apjl, 472, L81+

\bibitem[{{Iben} \& {Tutukov}(1985)}]{iben85}
{Iben}, I., Jr., \& {Tutukov}, A.~V. 1985, \apjs, 58, 661

\bibitem[{{Isern} et~al.(2010){Isern}, {Garc{\'{\i}}a--Berro}, {Althaus}, \&
  {C{\'o}rsico}}]{iser10}
{Isern}, J., {Garc{\'{\i}}a--Berro}, E., {Althaus}, L.~G., \& {C{\'o}rsico},
  A.~H. 2010, \aap, 512, A86+

\bibitem[{{Isern} et~al.(2008){Isern}, {Garc{\'{\i}}a--Berro}, {Torres}, \&
  {Catal{\'a}n}}]{iser08}
{Isern}, J., {Garc{\'{\i}}a--Berro}, E., {Torres}, S., \& {Catal{\'a}n}, S.
  2008, \apjl, 682, L109

\bibitem[{{Li} et~al.(2011){Li}, {Chornock}, {Leaman}, {Filippenko},
  {Poznanski}, {Wang}, {Ganeshalingam}, \& {Mannucci}}]{li11b}
{Li}, W., {Chornock}, R., {Leaman}, J., {Filippenko}, A.~V., {Poznanski}, D.,
  {Wang}, X., {Ganeshalingam}, M., \& {Mannucci}, F. 2011, \mnras, 412, 1473

\bibitem[{{Lor{\'e}n-Aguilar} et~al.(2009){Lor{\'e}n-Aguilar}, {Isern}, \&
  {Garc{\'{\i}}a--Berro}}]{lore09}
{Lor{\'e}n-Aguilar}, P., {Isern}, J., \& {Garc{\'{\i}}a--Berro}, E. 2009, \aap,
  500, 1193

\bibitem[{{Maoz} et~al.(2010){Maoz}, {Sharon}, \& {Gal-Yam}}]{maoz10}
{Maoz}, D., {Sharon}, K., \& {Gal-Yam}, A. 2010, \apj, 722, 1879

\bibitem[{{Motl} et~al.(2007){Motl}, {Frank}, {Tohline}, \& {D'Souza}}]{motl07}
{Motl}, P.~M., {Frank}, J., {Tohline}, J.~E., \& {D'Souza}, M.~C.~R. 2007,
  \apj, 670, 1314

\bibitem[{{Nomoto}(1982)}]{nomo82b}
{Nomoto}, K. 1982, \apj, 257, 780

\bibitem[{{Nomoto} \& {Kondo}(1991)}]{nomo91}
{Nomoto}, K., \& {Kondo}, Y. 1991, \apjl, 367, L19

\bibitem[{{Nomoto} et~al.(1985){Nomoto}, {Thielemann}, \& {Miyaji}}]{nomo85}
{Nomoto}, K., {Thielemann}, F.-K., \& {Miyaji}, S. 1985, \aap, 149, 239

\bibitem[{{Piersanti} et~al.(2003{\natexlab{a}}){Piersanti}, {Gagliardi},
  {Iben}, \& {Tornamb{\'e}}}]{pier03b}
{Piersanti}, L., {Gagliardi}, S., {Iben}, I., Jr., \& {Tornamb{\'e}}, A.
  2003{\natexlab{a}}, \apj, 598, 1229

\bibitem[{{Piersanti} et~al.(2003{\natexlab{b}}){Piersanti}, {Gagliardi},
  {Iben}, \& {Tornamb{\'e}}}]{pier03a}
--- 2003{\natexlab{b}}, \apj, 583, 885

\bibitem[{{Postnov} \& {Yungelson}(2006)}]{post06}
{Postnov}, K.~A., \& {Yungelson}, L.~R. 2006, Living Reviews in Relativity, 9,
  6

\bibitem[{{Rasio} \& {Shapiro}(1994)}]{rasi94}
{Rasio}, F.~A., \& {Shapiro}, S.~L. 1994, \apj, 432, 242

\bibitem[{{Ritossa} et~al.(1999){Ritossa}, {Garc{\'{\i}}a--Berro}, \&
  {Iben}}]{GB99}
{Ritossa}, C., {Garc{\'{\i}}a--Berro}, E., \& {Iben}, I., Jr. 1999, \apj, 515,
  381

\bibitem[{{Rosswog}(2007)}]{ross07}
{Rosswog}, S. 2007, \mnras, 376, L48

\bibitem[{{Saio} \& {Nomoto}(2004)}]{saio04}
{Saio}, H., \& {Nomoto}, K. 2004, \apj, 615, 444

\bibitem[{{Segretain} et~al.(1997){Segretain}, {Chabrier}, \&
  {Mochkovitch}}]{segretain97}
{Segretain}, L., {Chabrier}, G., \& {Mochkovitch}, R. 1997, \apj, 481, 355

\bibitem[{{Sim} et~al.(2010){Sim}, {R{\"o}pke}, {Hillebrandt}, {Kromer},
  {Pakmor}, {Fink}, {Ruiter}, \& {Seitenzahl}}]{sim10}
{Sim}, S.~A., {R{\"o}pke}, F.~K., {Hillebrandt}, W., {Kromer}, M., {Pakmor},
  R., {Fink}, M., {Ruiter}, A.~J., \& {Seitenzahl}, I.~R. 2010, \apjl, 714,
  L52. \eprint{1003.2917}

\bibitem[{{Sullivan} et~al.(2006){Sullivan}, {Le Borgne}, {Pritchet},
  {Hodsman}, {Neill}, {Howell}, {Carlberg}, {Astier}, {Aubourg}, {Balam},
  {Basa}, {Conley}, {Fabbro}, {Fouchez}, {Guy}, {Hook}, {Pain},
  {Palanque-Delabrouille}, {Perrett}, {Regnault}, {Rich}, {Taillet}, {Baumont},
  {Bronder}, {Ellis}, {Filiol}, {Lusset}, {Perlmutter}, {Ripoche}, \&
  {Tao}}]{sull06}
{Sullivan}, M., {Le Borgne}, D., {Pritchet}, C.~J., {Hodsman}, A., {Neill},
  J.~D., {Howell}, D.~A., {Carlberg}, R.~G., {Astier}, P., {Aubourg}, E.,
  {Balam}, D., {Basa}, S., {Conley}, A., {Fabbro}, S., {Fouchez}, D., {Guy},
  J., {Hook}, I., {Pain}, R., {Palanque-Delabrouille}, N., {Perrett}, K.,
  {Regnault}, N., {Rich}, J., {Taillet}, R., {Baumont}, S., {Bronder}, J.,
  {Ellis}, R.~S., {Filiol}, M., {Lusset}, V., {Perlmutter}, S., {Ripoche}, P.,
  \& {Tao}, C. 2006, \apj, 648, 868

\bibitem[{{Totani} et~al.(2008){Totani}, {Morokuma}, {Oda}, {Doi}, \&
  {Yasuda}}]{TEA08}
{Totani}, T., {Morokuma}, T., {Oda}, T., {Doi}, M., \& {Yasuda}, N. 2008,
  \pasj, 60, 1327. \eprint{0804.0909}

\bibitem[{{Webbink}(1984)}]{webb84}
{Webbink}, R.~F. 1984, \apj, 277, 355

\bibitem[{{Whelan} \& {Iben}(1973)}]{whel73}
{Whelan}, J., \& {Iben}, I., Jr. 1973, \apj, 186, 1007

\bibitem[{{Yoon} et~al.(2007){Yoon}, {Podsiadlowski}, \& {Rosswog}}]{yoon07}
{Yoon}, S., {Podsiadlowski}, P., \& {Rosswog}, S. 2007, \mnras, 380, 933

\end{thebibliography}

\end{document}